\documentclass[a4paperal]{i3m}

\usepackage{graphicx}
\usepackage{siunitx}
\usepackage{balance}

\usepackage{bm}
\usepackage{amsmath} 
\usepackage{mathtools} 
\usepackage[capitalize, nameinlink]{cleveref}
\usepackage{subcaption}

\catcode`\%=12\relax
\DeclareSIUnit[number-unit-product = ]\percent{
\catcode`\%=14\relax
\sisetup{detect-weight=true,detect-inline-weight=math}

\conference{emss}

\title{Simulation of Social Media-Driven Bubble Formation in Financial Markets using an Agent-Based Model with Hierarchical Influence Network}

\author[1]{Gonzalo Bohorquez}
\author[2,\authfn{1}]{John Cartlidge}

\affil[1]{School of Computer Science, University of Bristol, Bristol, UK}
\affil[2]{School of Engineering Mathematics and Technology, University of Bristol, Bristol, UK}

\authnote{\authfn{1}Corresponding author. Email address: john.cartlidge@bristol.ac.uk}

\runningauthor{Bohorquez and Cartlidge}

\confyear{2024}

\begin{document}

\begin{frontmatter}
\maketitle
\begin{abstract}
We propose that a tree-like hierarchical structure represents a simple and effective way to model the emergent behaviour of financial markets, especially markets where there exists a pronounced intersection between social media influences and investor behaviour. To explore this hypothesis, we introduce an agent-based model of financial markets, where trading agents are embedded in a hierarchical network of communities, and communities influence the strategies and opinions of traders. Empirical analysis of the model shows that its behaviour conforms to several stylized facts observed in real financial markets; and the model is able to realistically simulate the effects that social media-driven phenomena, such as echo chambers and pump-and-dump schemes, have on financial markets. 
\end{abstract}

\end{frontmatter}


\section{Introduction}
\label{sec:introduction}
In recent years, there has been a dramatic shift in the accessibility of financial markets. New {\em fintech} applications such as {\em Robinhood} and {\em eToro} have enabled retail investors to trade directly on traditional markets, without the need for expensive brokerage services; and the unprecedented disruption of new cryptocurrency assets, with their overall valuation rising from zero to more than one trillion dollars in less than 15 years, has driven crypto-speculation by millions of novice investors. At the same time, social media has become a ubiquitous presence in the lives of billions of people across the globe, which has influenced behaviours for both good and bad.

This shift in financial market participants and networks of influence has resulted in new causal drivers of market dynamics, leading to previously rare events becoming relatively commonplace (e.g., {\em GameStop short squeeze}, \cite{Klein-2022}). Additionally, the democratisation of markets has dramatically increased the opportunities for malicious actors to take advantage of the na\"ive through illicit practices such as {\em pump-and-dump schemes} \citep{li-2021}. 

{\bf Contribution}: In this paper, we introduce a new agent-based model of financial markets to capture this shift in dynamics. In particular, we extend the well-established model of Lux-Marchesi \citep{lux-marchesi-1999} by embedding agents on a hierarchical network to form {\em communities of influence}. These communities represent social media influence on behaviours. We are able to show that the model generates realistic behaviours under a variety of scenarios, and we are able to investigate the effects of social media influence, the effects of echo chambers, and markets most likely to be susceptible to pump-and-dump schemes. All code is available open source: {\href{https://github.com/gonzalo-bo/Lux-Hierarchy}{https://github.com/gonzalo-bo/Lux-Hierarchy}}.

The rest of this paper is organised as follows. In \cref{section:state-of-art}, we present a detailed background, review related works, and identify a research gap. In \cref{sec:method}, we detail our model implementation and describe a set of metrics that we use for evaluation. \cref{sec:results} presents model results, demonstrating realistic general behaviours (\cref{sec:results-stylised}) and realistic responses to particular scenarios (\cref{sec:results-scenarios}). We then discuss findings and their significance (\cref{sec:discuss}), and conclude by briefly illustrating limitations and opportunities for further work (\cref{sec:conc}).

\section{State of the art}
\label{section:state-of-art}

\subsection{Social media and financial markets}\label{sec:background-social-media}

\noindent
\citet{wang-2022} established a causal link between social media sentiment and same-day stock returns by analysing online messages from stock investment forum {\em EastMoney Guba}. This finding is supported by \citet{muller-2023}, who discovered the sentiment expressed in {\em X} (formerly {\em Twitter}) messages has a similar effect, and the influence on returns is stronger for ``risky, volatile'' assets with a market price that tends to deviate from the underlying fundamental value. 

Sentiment expressed through social media has also been shown to significantly impact market volatility, such that positive news on social platforms can lead to rapid price increases, while negative news can cause sudden declines. For instance, \citet{gilbert-2010} demonstrated a direct link between the sentiment on social media platforms and subsequent movements in the stock market, suggesting that increases in subjective expressions of anxiety could predict an increase in volatility for the S\&P 500 index. Additionally, \citet{jiao-2020} found that, whilst influence from traditional news media predicts a decrease in an asset’s volatility, influence from social media has the opposite effect. They present evidence suggesting that this distinction can most accurately be explained by an ``echo chamber'' model, where traders in social media are disproportionately affected by information that gets repeated (i.e., echoes) within the network. 

The {\em Echo Chamber Effect} refers to the idea that social media platforms can create environments where investors are exposed predominantly to opinions and information that reinforce their existing beliefs \citep{cinelli-2021}. Such environments can lead to overconfidence and exacerbate market anomalies, as investors might ignore contrary evidence or broader market signals. \citet{barber-2008} found that individual investors tend to trade more aggressively under the influence of overconfidence, which could be heightened by echo chambers in social media settings, leading to suboptimal trading decisions.

\citet{cookson-2022} provide evidence for the existence of the Echo Chamber Effect by analysing the data of more than 400,000 users on social network {\em StockTwits}. They reveal that users of this forum have a strong tendency to selectively expose themselves to information that reinforces their pre-existing beliefs, which leads to persistent disagreement within the forum and may help explain why assets subject to social media chatter display more price volatility. Furthermore, studies such as \citet{jiao-2020} and \citet{cookson-2022} suggest that the Echo Chamber Effect is more prevalent for positive opinions than for negative opinions; which leads to social media communities displaying disproportionately bullish tendencies (i.e., expectations that future prices will increase) regarding their most frequently discussed assets or asset classes.

Social media is also associated with the illicit practice of {\em pump-and-dump}, which is described by \citet{li-2021} as the act of artificially inflating (i.e., ``pump'') the price of an asset through exaggerated or entirely fabricated statements. Then, once the asset has been bought by unsuspecting buyers (drawn in by the hype), the schemers sell (i.e., ``dump'') their holdings at the elevated price, causing excess supply which precipitates a sudden collapse in market price and significant losses for the unwitting investors left holding the (essentially worthless) asset. \citet{lund-2022} suggests that pump-and-dump schemes are often facilitated by the reach and anonymity of social media, where it is easy to spread misleading information quickly to large and geographically dislocated audiences.

\subsection{Lux-Marchesi model of financial markets (LM99)}\label{sec:LM99}
\noindent
Lux-Marchesi---hereafter referred to as LM99---is a seminal agent-based model of financial markets \citep{lux-marchesi-1999}. Despite being 25 years old, its remarkable simplicity and versatility results in LM99 frequently being used as a foundation of contemporary agent-based models of financial markets \citep[e.g.,][]{alfarano-2011note, meine-2023}. 

LM99 contains a single financial asset with a fundamental value, $p_f$, representing the ``true'' price of an asset in an efficient market containing rational agents with perfect information. Two types of trading agent strategy populate the model: {\em fundamentalists} and {\em chartists}.  Fundamentalists make trading decisions based on the difference between the current market price and the current fundamental value, such that the asset is bought if market price is below the fundamental (i.e., when the asset is under-priced) and sold if the market price is above the fundamental (i.e., when the asset is over-priced). In this way, fundamentalists tend to drive the market price toward the fundamental value. In comparison, chartists make speculative trading decisions based on sentiment. Chartists fall into two categories: {\em optimists}, who are {\em bullish} and will always buy on the assumption that the market price will rise; and {\em pessimists}, who are {\em bearish} and will always sell on the assumption that the market price will fall. In this way, chartists tend to drive the market away from the fundamental value.

Importantly, agents in LM99 are adaptive and can switch between chartist and fundamentalist trading strategies based on the success, or profitability, of their current approach. LM99 also incorporates a mechanism for the diffusion of opinions among chartists, such that pessimists (optimists) are more likely to become optimists (pessimists) if lots of optimists (pessimists) are observed in the market. This simple mechanism enables the model to capture the psychological aspects of trading and the impact of collective sentiment. Together, these evolutionary mechanisms, where agents adapt strategy over time, allow the relatively simple model of LM99 to generate complex dynamics that realistically reproduce key aspects of real-world financial markets. For full details, see \cite{lux-marchesi-1999,lux-marchesi-2000}.

\subsection{Hierarchical models of social media networks}
\label{sec:hierarchical-models-lit}
\noindent
When attempting to model the interplay between social networks and financial markets, hierarchies repeatedly emerge as a simple yet effective way to represent several interesting phenomena. Modeling social network dynamics via hierarchies is a well-established practice that has proven to be quite powerful through several studies \citep{watts-2002,clauset-2008}. Furthermore, papers such as \citet{zhang-2017} and \citet{dang-2019} showcase the substantial improvements that hierarchical designs provide when it comes to modeling social media networks specifically.

\subsection{Hierarchical ABMs of financial markets}\label{sec:background-hiearchical-ABMs}
\noindent
While the apparent usefulness of hierarchical structures to model social networks is clear, there are very few papers exploring their use as a way to model communication between financial trading agents in an ABM. A systematic {\em Google Scholar} search for ``{\em (abm OR ``agent based model'') AND ``financial market'' AND (intitle:hierarchy OR intitle:hierarchical)}'' returns only ten unique results, and only two of these are pertinent to the current study. These are: \cite{alfarano-2011note}, who show that an ABM with a hierarchical-like structure leads to increased volatility in simulated markets; and \citet{meine-2023}, who reveal that a hierarchical structure leads to agents in simulated markets displaying a high degree of speculative behaviour. 
We therefore present the use of hierarchical networks to model social media influence in agent-based models of financial markets as a relatively underexplored gap in the research literature.

The model introduced by \citet{meine-2023}---hereafter referred to as MV23---extends the model of LM99 by embedding trading agents into a hierarchical network structure, such that each level in the network contains communities of {\em children} traders with mutual connection to a single {\em parent}. These communities represent traders interacting via online forums or social media groups, and the opinions (i.e., optimism/pessimism) and behaviours (i.e., chartist/fundamentalist) of community members affect the opinions and behaviours of others. Adding this network community structure enables the model of MV23 to demonstrate a greater variety of speculative {\em herding} behaviour than the underlying LM99 model, which it extends. 

\subsection{Research gap: social media-driven financial ABM}\label{sec:gap}

We have seen that social media networks have a significant effect on the behaviour of financial markets (\cref{sec:background-social-media}), and hierarchical network models can successfully model aspects of social media dynamics (\cref{sec:hierarchical-models-lit}). Combined, the two aforementioned ideas highlight the potential of applying hierarchical structures to models of social media communication between participants trading in a financial market. While some models, such as MV23, explore hierarchical structures as a way to model communication dynamics within financial markets (\cref{sec:background-hiearchical-ABMs}), we identified no models in the literature that are explicitly tailored to the context of social media. Therefore, we present hierarchical agent-based models of social-media influence on financial markets as a research gap in the literature.

In this paper, we address this research gap by introducing and exploring a hierarchical extension of LM99 (\cref{sec:LM99}). We initially considered generalizing MV23 to explore social media effects under different scenarios. However, we found that the implementation details of MV23 are difficult to interpret and the model's parameter configurations are not presented in \citet{meine-2023} \citep[for detailed analysis, see][]{gonzalo-MEng}. Therefore, while we acknowledge MV23 as a strong inspiration for the present work, we present an alternative model.

\section{Materials and methods}\label{sec:method}

In this section, we introduce a new agent-based model of financial markets, where trading agent opinions and behaviours are influenced by a hierarchical network of communities. We begin by detailing LM99, which forms the base of the model (Section~\ref{sec:LM99-eqs}), before describing the community network structure that we introduce as an extension (Section~\ref{sec:model-definition}). Finally, we define a series of metrics that will be used to evaluate the model (Section~\ref{sec:stylized}). 

\begin{table*}[tb]
    \caption{Parameter settings. Top: Values taken from \citet[][p.692]{lux-marchesi-2000}. Bottom: Parameter settings of hierarchical influence network.}
    \label{tab:params}
    \centering
    \begin{tabularx}{\linewidth}{c S[table-format=1.2] S[table-format=1.2] S[table-format=1.2] X}
    \toprule
         \textbf{} & \textbf{II} & \textbf{III} & \textbf{IV} & \textbf{Description} \\ 
    \midrule
         $\alpha _{2}$ & 0.25 & 0.25 & 0.2 & How much chartists are influenced by changes in the asset price\\ 
         $\alpha _{3}$ & 1& 0.75&1 & How much traders are influenced by a role's profit \\ 
         $v_{1}$ & 4& 0.5&2 & How often pessimists try to become optimists \& vice versa\\
         $v_{2}$ & 1& 0.5&0.6 & How often fundamentalists try to become chartists \& vice versa\\  
         $\beta$ & 4& 2&4& How often the market price changes\\ 
         $r$ & 0.004& 0.004&0.004 & Dividends paid by the asset\\
         $R$ & 0.0004& 0.0004&0.0004 & Returns from alternative investments\\ 
         $s$ & 0.75& 0.75&0.75 & Factor by which a fundamentalist's profit is reduced (i.e., discount factor)\\
         $p_{f}$ & 10& 10&10& Fundamental value\\ 
         $\sigma$ & 0& 0&0 & Magnitude of fundamental value fluctuations\\ 
         $\mu$ & 0.1& 0.1&0.05 & Noise when price changes due to excess demand/supply\\ 
         $\gamma$ & 0.01& 0.02&0.01& How strongly fundamentalists react to deviations from the fundamental price\\ 
         $t_{c}$ & 0.015& 0.02&0.01& How much of the asset is bought or sold (by optimists or pessimists respectively)\\
         $\delta t$ & 0.01& 0.01&0.01& Time-step/interval. The simulation runs for $\frac{1}{\delta t}$ time-steps per unit time \\ 
         $\delta t^\prime$& 0.002& 0.002&0.002 & Compare the current price and the price from $\delta t^\prime$ time-steps ago to determine how fast the price is changing\\ 
         \midrule
          $L$& 5& 5&5 & Number of levels in the hierarchy  \\ 
          $k$& 5& 5&5& Number of children per community  \\ 
          $b$& 1.8& 2.25&2.4& Strength of hierarchy (community) influence on optimist $\leftrightarrow$ pessimist transition  \\ 
          $\phi$& 0.5& 0.5&0.5& Efficiency of information diffusion within the hierarchy \\ 
          $\omega$& 1& 1&1& Optimist child's influence on its parent  \\ 
          $\upsilon$& 1& 1&1& Pessimist child's influence on its parent \\ 
    \bottomrule
    \end{tabularx}
    \begin{tablenotes}
	\item \textsuperscript{*}\citet[][p.692]{lux-marchesi-2000} parameters $N$ (number of traders) and $\alpha_1$ (influence of chartists on chartists) are replaced by network hierarchy; see equations \eqref{eqn:C-N} - \eqref{eq:forward-pass}.
    \end{tablenotes}
\end{table*}

\subsection{Model core: LM99}
\label{sec:LM99-eqs}
The core of the agent-based model closely follows LM99, introduced in \cite{lux-marchesi-1999}. We refer the reader to \cref{tab:params} for a description of model parameters and configuration values.

As described in Section~\ref{sec:LM99}, LM99 contains two agent types: {\em fundamentalists} and {\em chartists}, with chartists taking one of two roles: {\em optimist} or {\em pessimist}. The model contains a single asset and agents can perform one of two actions: buy the asset; or sell the asset. Excess profit is used to measure the opportunity cost inherent to these two actions; i.e., if an agent bought the asset, its excess profit is how much it made from the purchase minus how much it would have made from not buying the asset, and vice-versa for the excess profit if it decided to sell. 
We calculate the excess profits for fundamentalists, $EP_{f}$, excess profits for optimists, $EP_{+}$, and excess profits for pessimists, $EP_{-}$, as:
\begin{equation}
\label{eq:ep}
    \begin{gathered}
        EP_{f} = s \left|\frac{p_{f} - p}{p}\right|, \quad \ 
        EP_{+} = \left( r + \frac{\dot{p}}{v_{2}} \right)/p - R,      \vspace{0.2cm}  \\  EP_{-} = R - \left( r + \frac{\dot{p}}{v_{2}} \right)/p
    \end{gathered}
\end{equation}
\noindent where $p$ is the current market price, $p_{f}$ is the current fundamental price, and $\dot{p}$ is the price trend, calculated by comparing the current price with the price from $\delta_{t}'$ time-steps ago:
\begin{equation}
\label{eq:pdot}
\dot{p} = \left(p_{t} - p_{t - \delta_{t}'}\right) / \delta_{t}'
\end{equation} 

At any given time-step $t$, traders have a probability of switching from one role to another. This probability depends on a set of transition pressures. As defined in equation \eqref{eq:U1}, $U_{21}$ represents the pressure exerted by the $optimist \leftrightarrow fundamentalist$ transition, with positive values of $U_{21}$ meaning optimists feel pressure to become fundamentalists (vice-versa for negative values). Similarly, $U_{22}$ and $U_{1}$ represent the pressure exerted by the $pessimist \leftrightarrow fundamentalist$ and the $optimist \leftrightarrow pessimist$ transitions, respectively:
\begin{equation}\label{eq:U1}
    \begin{gathered}
        U_{21} = \alpha_{3}\left(EP_{f} - EP_{+}\right), \quad \ 
        U_{22} = \alpha_{3}\left(EP_{f} - EP_{-}\right),      \vspace{0.2cm}  \\  U_{1} = \alpha_{1}\left(\frac{n_p - n_o}{n_p + n_o}\right) + \alpha_{2}\frac{\dot{p}}{v_{1}}
    \end{gathered}
\end{equation}

\noindent where $n_o$ and $n_p$ represent the number of optimists and pessimists in the network, respectively.

Furthermore, as in real markets, the trader's actions affect the market price of the asset being traded. At any given time-step, the probability that the price rises, $\pi_{\uparrow p} $, or falls, $\pi_{\downarrow p}$, is determined by the total demand in the market:
\begin{equation}
\label{eq:price-update}
\begin{aligned}
    \pi_{\uparrow p} & = max[0, \beta(ED_{c} + ED_{f} + \mu)] \\
    \pi_{\downarrow p} & = min[0, -\beta(ED_{c} + ED_{f} + \mu)]
\end{aligned}
\end{equation}

$ED_{c}$ represents the total demand from chartists and is determined by the total number of optimists, $n_{o}$, and pessimists, $n_{p}$. $ED_{f}$ represents the total demand exerted by fundamentalists and is determined by the number of fundamentalists, $n_{f}$, as well as whether fundamentalists are buying or selling the asset. $ED_{c}$ and $ED_{f}$ are defined as:
\begin{equation}\label{eq:total-demand}
\begin{aligned}
ED_{c} & = (n_{o} - n_{p}) * t_{c} \\
ED_{f} & = n_{f} * \gamma * (p_{f} - p)
\end{aligned}
\end{equation}

Finally, equations \eqref{eq:transition-probs-op}, \eqref{eq:transition-probs-of} and \eqref{eq:transition-probs-pf} outline the transition probabilities between trader states:
\begin{equation}
\label{eq:transition-probs-op}
    \begin{gathered}
       \pi_{o \rightarrow p} = v_{1} \cdot \frac{n_c}{N} \cdot e^{-U_{1}} \cdot \delta_{t}, \quad \ 
        \pi_{p \rightarrow o} = v_{1} \cdot \frac{n_c}{N} \cdot e^{U_{1}} \cdot \delta_{t}
    \end{gathered}
\end{equation}
\vspace{-6mm}
\begin{equation}
\label{eq:transition-probs-of}
    \begin{gathered}
        \pi_{o \rightarrow f} = v_{2} \cdot \frac{n_o}{N} \cdot e^{-U_{21}} \cdot \delta_{t} , \quad \ 
        \pi_{f \rightarrow o} = v_{2} \cdot \frac{n_f}{N} \cdot e^{U_{21}} \cdot \delta_{t}
    \end{gathered}
\end{equation}
\vspace{-6mm}
\begin{equation}
\label{eq:transition-probs-pf}
    \begin{gathered}
        \pi_{p \rightarrow f} = v_{2} \cdot \frac{n_p}{N} \cdot e^{-U_{22}} \cdot \delta_{t}  , \quad \ 
        \pi_{f \rightarrow p} = v_{2} \cdot \frac{n_f}{N} \cdot e^{U_{22}} \cdot \delta_{t}
    \end{gathered}
\end{equation}
\noindent where $\pi_{a \rightarrow b}$ represents the probability that a trader of type $a$ becomes a trader of type $b$ at any given time-step, and $n_c$ represents the total number of chartists in the market. 

\subsection{Model extension: hierarchical network}\label{sec:model-definition}

We extend LM99 by embedding agents within a hierarchical network structure, which allows us to model a trading agent's opinions and behaviours being influenced by other members of their local community and wider community-of-communities. The network contains two types of node: {\em community} and {\em trading agent}. Trading agents exist within the bottom level (leaf nodes) of the hierarchy, while all higher-level nodes are designated as communities. 

We assume that the network hierarchy has $L$ levels and every non-leaf node has exactly $k$ children. Then the total number of trading agents, $N_t$, is equal to the number of leaf nodes, and the total number of communities, $N_c$, is equal to the number of non-leaf nodes, i.e.:
\begin{align}
\begin{split}
\label{eqn:C-N}
    N_t= k^{L-1} ,
    \quad \ N_c = \frac{k^L - 1}{L - 1} - N_t
\end{split}
\end{align}

Unlike the base model of LM99, where agents have perfect knowledge of the distribution of opinions across the whole market, in our extended model, chartists only have access to information that is transmitted through the network structure. As such, the $optimist \leftrightarrow pessimist$ transition equation \eqref{eq:U1} is replaced by equation \eqref{eq:optimist-pessimist-transition-modified}:
\begin{equation}\label{eq:optimist-pessimist-transition-modified}
U_{1} = b \left( \frac{C_o - C_p}{C_o + C_p} \right) + \alpha_{2}\frac{\dot{p}}{v_{1}}
\end{equation} 
\noindent where $b$ represents {\em hierarchy Strength}, and $C_o$, $C_p$ represent the proportion of optimists and pessimists in the local community, respectively. 

Community nodes do not store a discrete state but rather a vector: [$o$, $p$, $f$], where $o$, $p$, and $f$ are positive, continuous variables representing the optimist, pessimist, and fundamentalist states respectively.

Additionally, by letting traders behave like a homogeneous community (i.e., a community only populated by one child type), equation \eqref{eq:community-nodes} simplifies modelling the interaction between communities and traders:
\begin{equation}\label{eq:community-nodes}
    \begin{aligned}
        Optimist & = [\omega, 0, 0] \\
        Pessimist & = [0, \upsilon, 0] \\
        Fundamentalist & = [0, 0, 1]
    \end{aligned}
\end{equation}
\noindent where $\omega$ and $\upsilon$ represent the influence of optimist and pessimist children, respectively.

Community nodes update their state in two distinct phases:
\begin{enumerate}
    \item \textbf{Backward pass}: for each child, $u$, a community's state is updated according to equation \eqref{eq:backward-pass}. In other words, each community state, $C$, is determined by the average of its $k$ children's states:
\begin{equation}
\label{eq:backward-pass}
C_{[o,\ p,\ f]} = \frac{1}{k} \sum u_{[o,\ p,\ f]}
\end{equation}
\item \textbf{Forward pass}: following equation \eqref{eq:forward-pass}, each community state, $C$, is updated to account for the state of their respective parent community, $Q$:
\begin{equation}
\label{eq:forward-pass}
C'_{[o,\ p,\ f]} = \frac{1}{2} C_{[o,\ p,\ f]} + \phi \cdot Q_{[o,\ p,\ f]}
\end{equation}
\end{enumerate}
\noindent where $\phi$ represents the \textit{network efficiency of information diffusion}. As $\phi \to\infty$, each trader is equally affected by the averaged opinions of all other traders in the network, while lower values of $\phi$ imply that traders are more strongly affected by opinions in the local community, and $\phi = 0$ represents insular traders that are not affected by the opinions of any other trader, regardless of network distance.

At each time-step, the backward pass is followed by the forward pass, which is then followed by traders switching their state according to the model's transition pressures. Finally, following the LM99 paradigm defined in equation \eqref{eq:price-update}, the market price is updated through the interaction of trading agents buying and selling.

The full set of model parameters and their descriptions are listed in \cref{tab:params}. For the base LM99 parameters, we follow the three standard parameter sets presented in \citet[][p.692]{lux-marchesi-2000}. Note that number of traders, $N$, is replaced by $N_t$ from equation \eqref{eqn:C-N} and influence of chartists, $\alpha_1$, is replaced by $b$ as described in equation \eqref{eq:optimist-pessimist-transition-modified}. 

Network parameters are presented at the bottom of \cref{tab:params}. We fix hierarchy levels $L=5$ and children per community $k=5$, giving $N_t=625$ trading agents in total as per equation \eqref{eqn:C-N}. To preserve model symmetry, the influence of optimists $\omega=1$ and pessimists $\upsilon=1$ are fixed. Efficiency of information diffusion $\phi=0.5$ is fixed to balance the relative strength of influence of child and parent communities in equation \eqref{eq:forward-pass}. Finally, $b$ controls the strength of hierarchy influence on the $optimist \leftrightarrow pessimist$ transition, as defined in equation \eqref{eq:optimist-pessimist-transition-modified}.

Note that, while there is no toggle to quickly reduce the hierarchical model into the base LM99 model, some specific parameter arrangements will achieve this purpose, e.g., setting $b = 0$ in equation \eqref{eq:optimist-pessimist-transition-modified} is equivalent to LM99 with $\alpha_1 = 0$, and as $\phi \to\infty$ in equation \eqref{eq:forward-pass} the model approaches LM99 with $\alpha_1 = b$.

\subsection{Model evaluation: stylized facts}
\label{sec:stylized}

\noindent
To validate the {\em realism} of financial market simulations, it is common to compare the behaviour of the simulation model with the behaviour of real financial markets. Although real financial markets exhibit complex dynamics, there are broad patterns that consistently emerge. These so-called {\em stylized facts} can be used as a benchmark to validate simulation models, i.e., if the simulation reproduces the stylized facts of real markets, then we can consider that the simulation behaves in a somewhat realistic fashion. In this work, we consider four stylized facts of financial markets, and we describe a metric for detecting financial bubbles.

\subsubsection{Fat tails of returns}
\label{section:fat-tails-method}
The {\em fat tails} phenomenon is a stylized fact observed for an asset's (logarithmic) returns, indicating that large deviations from the mean are more common than would be expected under a normal distribution. Fat tails are usually represented through leptokurtic distributions; however, as explained in \citet{lux-marchesi-2000}, kurtosis is a somewhat ambiguous concept, and it is not entirely clear how to compare the kurtosis statistics obtained for various time series. 
For this reason, rather than measuring kurtosis, it is standard practice to measure fat-tailedness by assuming that the tails of the returns decay according to a Pareto distribution: $1 - ax^{-\alpha}$.

Equation \eqref{eq:hill} details how to estimate $\alpha$ for a tail of size $m$ out of a total of $n$ observations \citep{hill-1975}:
\begin{equation}
\label{eq:hill}
    \alpha_H = \frac{1}{\frac{1}{m} \sum_{i=1}^{m} \left[ \ln (R_{n - i - 1}) - \ln (R_{n - m}) \right]}
\end{equation}

\noindent
Following standard practice, we will present values for the $2.5\%$, $5\%$, and $10\%$ right-tails of the absolute returns.
\\ \\
\noindent \textit{{\bf Fact}: The absolute returns of most financial assets present a tail decay such that $2 \leq \alpha \leq 6$, with lower values of $\alpha$ representing fatter tails \citep{cont-2001}}.

\subsubsection{Volatility clustering}
\noindent

\noindent
The volatility of an asset, $\sigma$, is generally measured as the standard deviation of market price, defined as:
\begin{equation}
\label{eq:sigma}
    \sigma = \sqrt{\frac{\sum (p_i - \mu)^2}{N}}
\end{equation}
\noindent where $\mu$ is mean market price and $p_i$ is market price at time-step $i$. 

In simulation, but not in real markets, we can also calculate the deviation of market price from the fundamental value, $F_\sigma$, defined as: 
\begin{equation}
\label{eq:fundamental}
    F_\sigma = \sqrt{\frac{\sum (p_i - f_i)^2}{N}} - \sqrt{\frac{\sum (p_i - \mu)^2}{N}}
\end{equation}
\noindent where $f_i$ is the fundamental value at time-step $i$.

{\em Volatility clustering} is the observation that asset returns have a tendency to cluster based on their magnitude/significance. That is, large changes tend to be followed by large changes; while small changes tend to be followed by small changes. 
To evaluate volatility clustering, we measure the auto-correlation function (ACF) for the absolute and squared returns when $\tau = 10$ and $T = 70$. We compute ACF by importing the \texttt{acf} function from \texttt{statsmodels.api.tsf}.
\\ \\
\textit{{\bf Fact}: We expect ACF value to be larger than 0, indicating that, for all time-steps, there is a correlation between the current returns and returns from $10$ time-steps ago.}

\subsubsection{Return to gaussianity}
The {\em return to gaussianity} or {\em aggregational gaussianity} phenomenon refers to the expectation that the distribution of the asset's (logarithmic) returns should resemble a normal distribution for large values of $T$. To evaluate this, we measure the kurtosis of returns with $T = 1$, $T = 10$, and $T = 50$. 

To compute the excess kurtosis of the returns, we use the \texttt{norm.pdf} function from \texttt{scipy.stats} to fit the returns to a normal distribution, after which we use the \texttt{pandas.kurt} function to compute the distribution's kurtosis.
\\ \\
\textit{{\bf Fact}: We expect excess kurtosis to be large when $T=0$, and approach zero as $T$ becomes large.}

\subsubsection{Slow decay of autocorrelation}
This stylized fact is related to the observation that the autocorrelation in the absolute returns of an asset tends to decay slowly, following a power law such that $f(x) = at^{-\beta}$.

To evaluate the slow decay of the ACF, we measure the decay of absolute returns with $T=70$ as a function of $\tau$. This value is calculated by using the \texttt{curve\_fit} function of \texttt{scipy.optimize} library to fit an exponential decay function $f(x) = at^{-\beta}$ to the ACF of the absolute returns. 
\\ \\
\textit{{\bf Fact}: When calculated through GARCH-type models, real markets exhibit $\beta \in [0.2, 0.4]$ \citep{cont-2001}. However, using our methodology, we expect values of $\beta \in [0.6, 1]$.}

\begin{table}
    \caption{SADF and GSADF critical values at significance levels 90-100\%. }
    \label{tab:critical-values}
    \begin{tabularx}{\linewidth}{X S[table-format=1.2] S[table-format=1.2] S[table-format=1.2]  
    S[table-format=1.2] S[table-format=1.2] S[table-format=1.2]}

    \toprule
          & \multicolumn{3}{c}{Asymptotic SADF}  & \multicolumn{3}{c}{Finite-sample GSADF\textsuperscript{*}}\\
          \cmidrule(lr){2-4} \cmidrule(l){5-7}
          \textbf{$T$}&\textbf{90\%}&\textbf{95\%}&\textbf{100\%}  &\textbf{90\%}&\textbf{95\%}&\textbf{100\%}\\ 
    \midrule

          $\bm{100}$ & 1.10& 1.37&1.88 &     1.65& 2.00&2.57  \\ 
          $\bm{200}$ & 1.12& 1.41&2.03 &     1.84& 2.08&2.70\\ 
          $\bm{400}$ & 1.20& 1.49&2.07 &     1.92& 2.20&2.80\\ 
          $\bm{800}$ & 1.21& 1.51&2.06 &     2.10& 2.34&2.79\\ 
          $\bm{1600}$ & 1.23& 1.51&2.06 &    2.19& 2.41&2.87\\ 
    \bottomrule
    \end{tabularx}
    \begin{tablenotes}
        \item \textsuperscript{*}GSADF $(T,r_0)$ values: (100,0.190), (200,0.137), (400,0.100), (800,0.074), (1600,0.055).
    \end{tablenotes}
\end{table}

\subsubsection{Financial bubbles \& explosive behaviour}
{\em Explosive behaviour} in an asset price describes a rapid and anomalous increase. A related concept is a {\em bubble}, which also describes a rapid and anomalous increase in market price. However, to be classified as a bubble, it is necessary for the price increase to be driven by speculation rather than an increase in fundamental value; whereas explosive behaviour covers both scenarios. 

To detect financial market bubbles, we use the simple and well-documented GSADF test \citep{phillips-2015}. 
For the GSADF method with finite samples, we derive our significance level and minimum window size, $r_0$, values from \citet{phillips-2011}; see \cref{tab:critical-values}. 
We also use the PWY procedure to detect {\em explosive periods} \citep{phillips-2015}, and follow the methodology of \citet{phillips-2010} to identify significant bubbles through the asymptotic SADF critical values (see \cref{tab:critical-values}).

\section{Results}\label{sec:results}

In this section we first show that the model produces realistic dynamics (Section~\ref{sec:results-stylised}). We then use the model to explore real-world scenarios of social media influence, echo chambers, and pump-and-dump schemes (Section~\ref{sec:results-scenarios}).

\subsection{Reproducing stylized facts of financial markets}\label{sec:results-stylised}
\begin{figure*}[tbh]
\captionsetup[subfigure]{justification=centering}
\centering
\begin{subfigure}{.35\textwidth}
  \includegraphics[width=.9\linewidth]{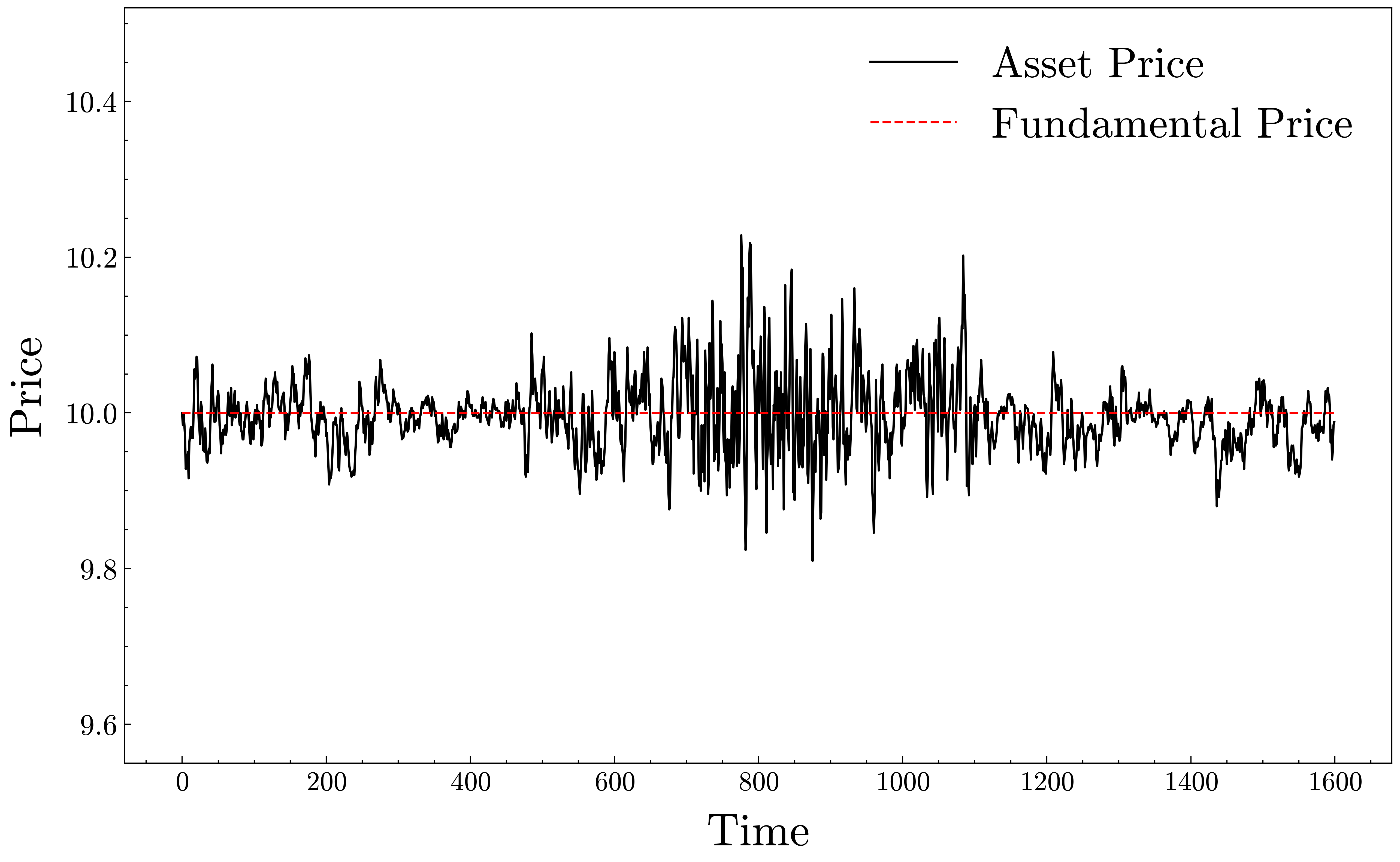}
  \caption{b = 0} 
  \label{fig:sample-b0}
\end{subfigure}%
\begin{subfigure}{.35\textwidth}
  \includegraphics[width=.9\linewidth]{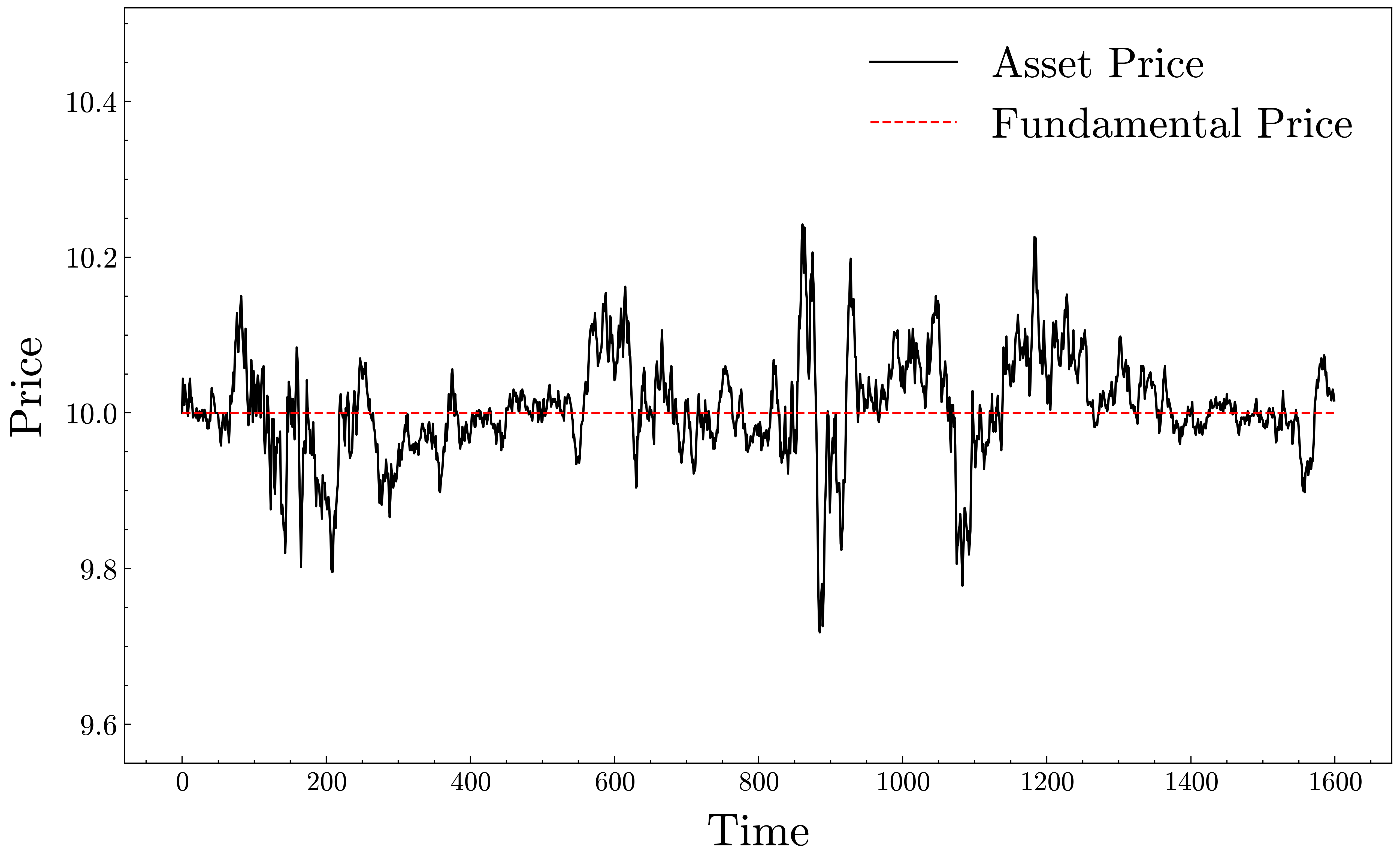}
  \caption{b = 1}
  \label{fig:sample-b1}
\end{subfigure}%
\begin{subfigure}{.35\textwidth}
  \includegraphics[width=.9\linewidth]{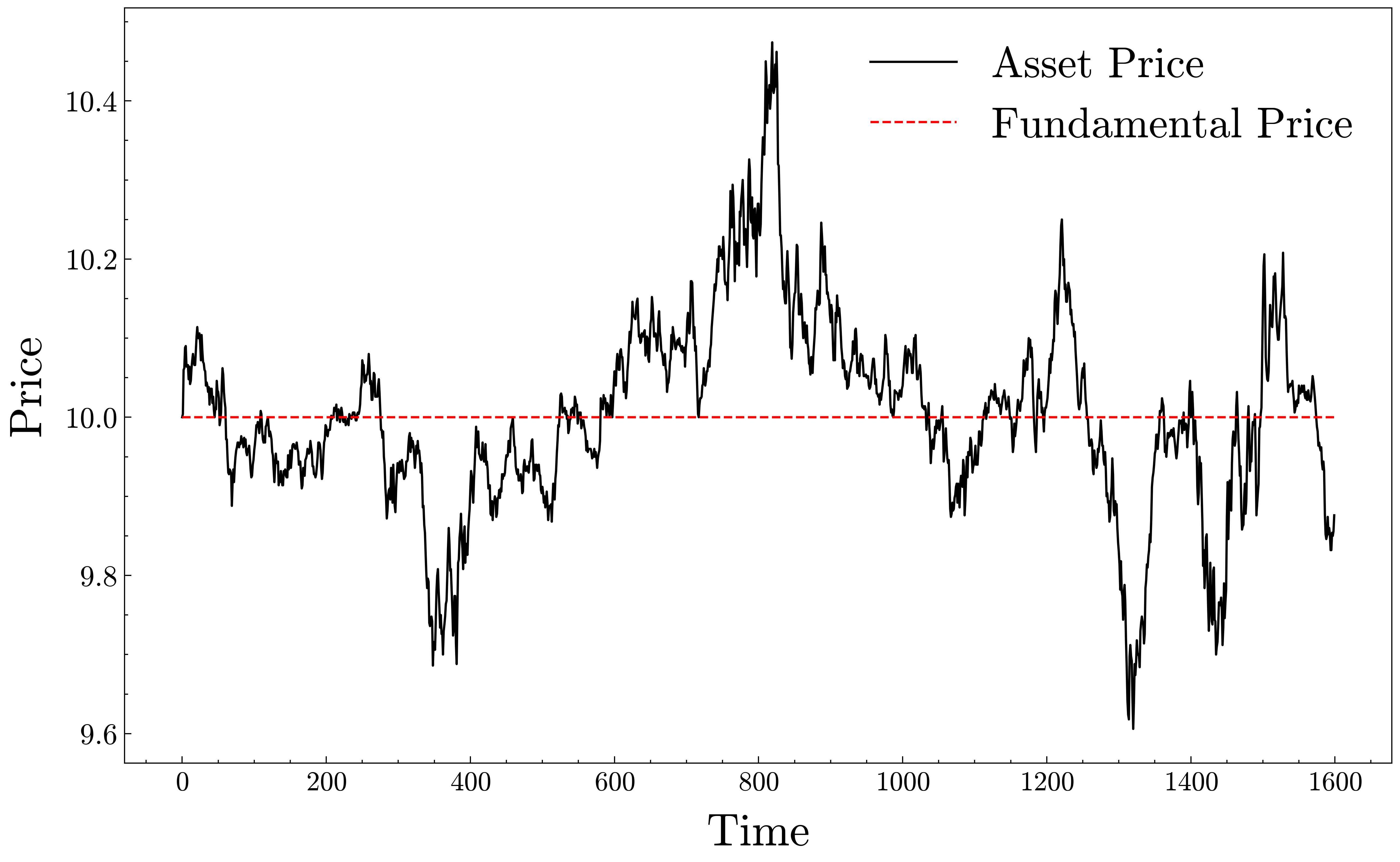}
  \caption{b = 2}
  \label{fig:sample-b2}
\end{subfigure}%
\caption{Sample runs showing how price volatility and fundamental deviation increases with hierarchy strength, $b$.}
\label{fig:sample-b}
\end{figure*}

\noindent
\cref{fig:sample-b} shows sample runs with fixed fundamental value, $p_f=10$, under increasing values of $b$ (refer to equation \eqref{eq:optimist-pessimist-transition-modified}). We see that price movements increase in magnitude as the hierarchy's effect becomes more pronounced, with the market price displaying sustained and significant deviation from the fundamental price when $b$ is large. 

As shown \cref{tab:volatility-results}, more rigorous analysis reveals that: for all parameter sets, an increase in $b$ is associated with a significant increase in price volatility and a significant increase in the likelihood of explosive periods.

\begin{figure}[tb]
\centering
  \includegraphics[width=0.9\linewidth]{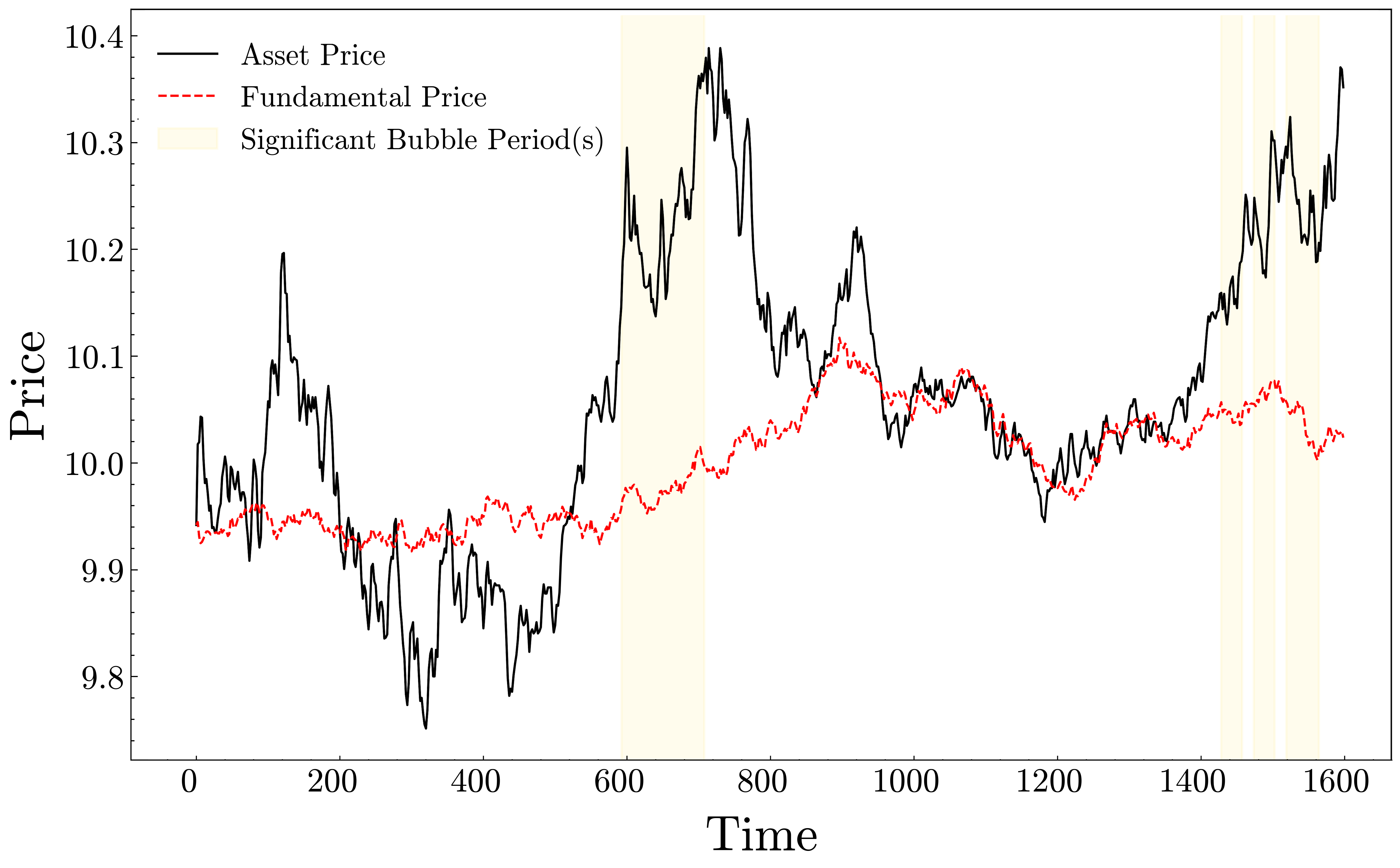}
\caption{PWY procedure on hierarchical model identifies explosive periods (shaded yellow) that are prolonged (left), and clustered in time (right).}
\label{fig:pwy}
\end{figure}

\begin{figure}[tb]
\centering
  \includegraphics[width=0.9\linewidth]{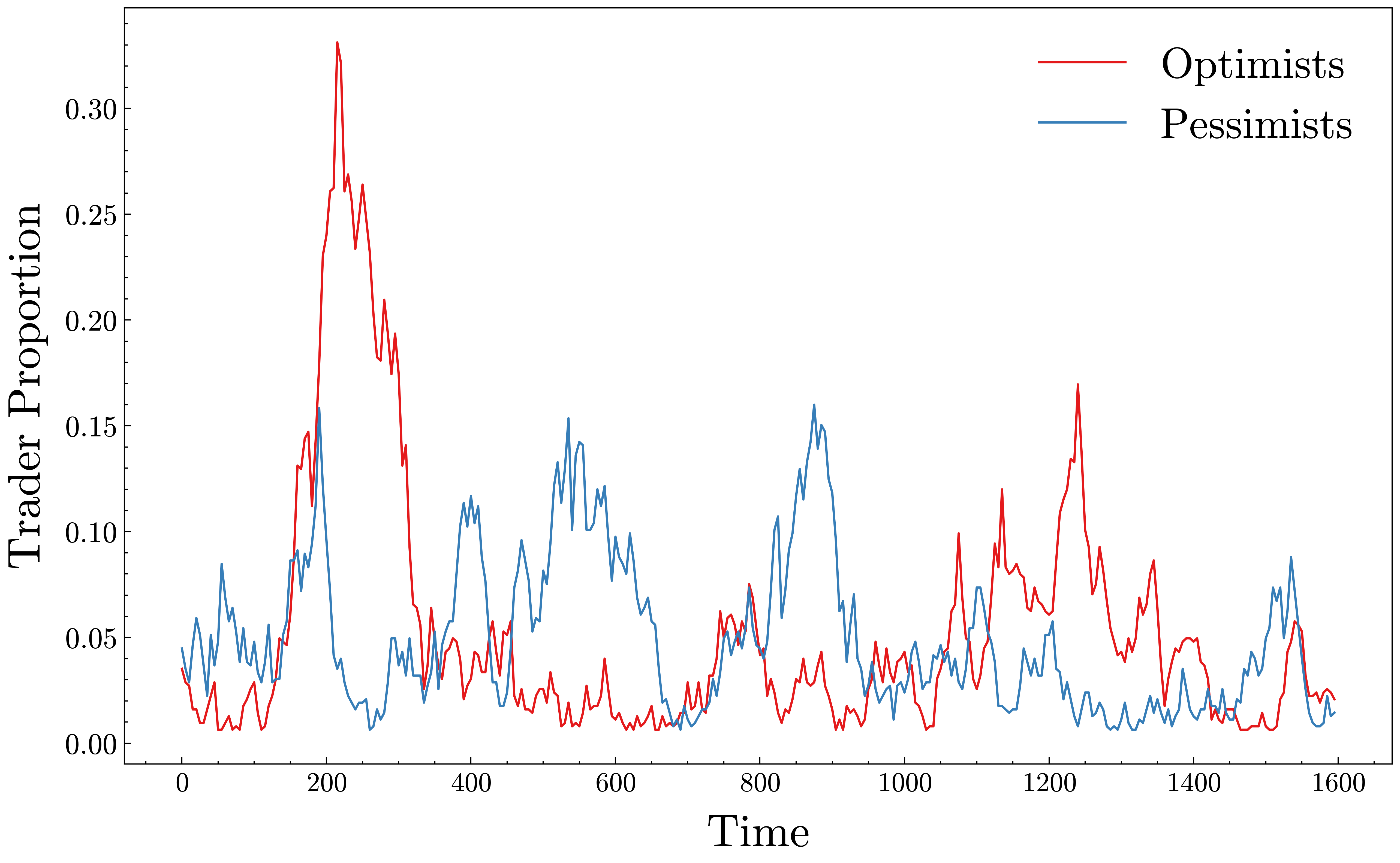}
  \caption{Opinion distributions within the chartist population reveals strong herding tendencies 
  (at T=175-300; 450-700; 1175-1350).}
\vspace{0.1cm}
\label{fig:hierarchy-emotion-pset3}
\end{figure}

Furthermore, by looking at individual runs (e.g., see \cref{fig:pwy}), we can see that instances of explosive behaviour are often sustained in time or occur in clustered bursts. This suggests that bubbles are driven by endogenous forces rather than occurring by random chance. Indeed, as shown in \cref{fig:hierarchy-emotion-pset3}, we see that the model leads to {\em herding behaviour} of chartists, such that traders collectively swing from optimists to pessimists and back again over time. This behaviour is not present in the core model of LM99 and results directly from the hierarchical influence network.

Finally, \cref{tab:stylized-standard} shows that all metrics fall within expected bounds, demonstrating that the simulation model is capable of reproducing all stylized facts of real markets introduced in \cref{sec:stylized}.

\begin{table}[tb]
 \caption{Mean volatility and proportion of runs with explosive periods, showing both measures tend to increase with hierarchy strength, $b$.}
 \begin{tabularx}{\linewidth}{S S S[table-format=1.1] S[table-format=1.1] S[table-format=1.1] S S[table-format=1.0] S[table-format=1.0] S[table-format=1.0]}
		\toprule
  {} & {} & \multicolumn{3}{c}{Volatility} & {} & \multicolumn{3}{c}{Explosive Periods} \\  \cmidrule(lr){3-5}\cmidrule(lr){7-9}
		{\textit{b}} & & {\textbf{Set II}} & {\textbf{Set III}} & {\textbf{Set IV}} & &{\textbf{Set II}} & {\textbf{Set III}} & {\textbf{Set IV}}\\
		\midrule
    \bfseries 0.0 & & 4.7 & 1.6 & 3.9  & & \SI{0}{\percent} & \SI{4}{\percent} & \SI{0}{\percent} \\
    \bfseries 0.1 & & 5.9 & 1.7 & 3.6 & & \SI{2}{\percent} & \SI{4}{\percent} & \SI{0}{\percent} \\
    \bfseries 0.25 & & 5.5 & 2.2 & 4.0 & & \SI{6}{\percent} & \SI{8}{\percent} & \SI{6}{\percent} \\
    \bfseries 0.5 & & 5.6 & 2.1 & 4.5 & & \SI{4}{\percent} & \SI{12}{\percent} & \SI{6}{\percent} \\
    \bfseries 1.0 & & 6.7 & 2.8 & 5.7 & & \SI{2}{\percent} & \SI{14}{\percent} & \SI{15}{\percent}\\
    \bfseries 2.0 & & 9.1 & 4.4 & 8.9 & & \SI{44}{\percent} & \SI{34}{\percent} & \SI{22}{\percent}\\
		\bottomrule
	\end{tabularx}
\begin{tablenotes}
	\item \textsuperscript{*}Cells show mean value of $50$ i.i.d. simulation trials, with $8 * 10^4$ time-steps.
        \item \textsuperscript{*}Volatility values are scaled by a factor of $10 ^{-2}$.
        \item \textsuperscript{*}Explosive behaviour is determined via the GSADF test at the 90\% significance level.
	\end{tablenotes}
 \label{tab:volatility-results}
\end{table}

\begin{table}[tb]
 \caption{Stylized facts reproduced by the model.}
 \label{tab:stylized-standard}
	\begin{tabularx}{\linewidth}{X S[table-format=1.2,table-column-width=0.6cm] S[table-format=1.2,table-column-width=0.6cm] S[table-format=1.2,table-column-width=0.6cm]}
		\toprule
		{} & {\textbf{Set II}} & {\textbf{Set III}} & {\textbf{Set IV}}\\
		\midrule
  \vspace{-0.2cm}
    \textit{Fat tails of (log) returns ($2\leq\ x\leq6$)}\\
    \cmidrule{1-4}
    \textbf{Tail Decay (2.5\%)} & 4.72 & 5.22 & 5.16\\
    \textbf{Tail Decay (5\%)} & 3.48 &  4.21 & 3.82 \\
    \textbf{Tail Decay (10\%)} & 2.72 & 3.24 & 2.83 \\
    \midrule
    \vspace{-0.2cm}
    \textit{Return to Gaussianity ($\lim_{T \to \infty} x = 0$)}\\
    \cmidrule{1-4}
    \textbf{Excess Kurtosis ($\bm{T = 1}$)} & 1.79 & 1.15 & 1.35\\
    \textbf{Excess Kurtosis ($\bm{T = 10}$)} & 1.72 & 1.12 & 1.22\\
    \textbf{Excess Kurtosis ($\bm{T = 50}$)} & 0.37 & 0.02 & 0.16\\ 
    \midrule
    \vspace{-0.2cm}
    \textit{Volatility clustering of returns ($x>0$)}\\
    \cmidrule{1-4}
    \textbf{Log Autocorrelation ($\bm{\tau = 10, T=70}$)} & 0.29 & 0.52 & 0.45\\
    \textbf{Squared Autocorrelation ($\bm{\tau = 10, T=70}$)} & 0.4 & 0.49 & 0.40\\
    
    \midrule
    \vspace{-0.2cm}
    \textit{Slow decay of returns autocorrelation ($x<1$)}\\
    \cmidrule{1-4}
    \textbf{Absolute Autocorrelation Decay} & 0.77 & 0.68 & 0.67\\
    \bottomrule
    \end{tabularx}
 \begin{tablenotes}
             \item \textsuperscript{*}Cells show mean value of $50$ i.i.d. simulation trials, with $4 * 10^4$ time-steps.
	\end{tablenotes}
\end{table}

\subsection{Scenario testing}\label{sec:results-scenarios}

In the previous section, we validated the model by showing that it produces realistic financial market dynamics. In this section, we {\em scenario test} the model to see how it behaves under specific situations of interest. 

\subsubsection{Social media effect on volatility and bubbles}
\label{section:efficient}

\noindent
Probably the most salient distinction between social media networks and their traditional counterparts is the speed and efficiency with which information can be distributed across the former medium. Here, we note that higher values of $\phi$ in equation \eqref{eq:forward-pass} represent less network decay, equivalent to higher network efficiency. Thus, we take larger values of $\phi$ to represent social media, while smaller values of $\phi$ represent slower, more conventional modes of information diffusion.

\begin{figure}[tb]
\centering
  \includegraphics[width=0.9\linewidth]{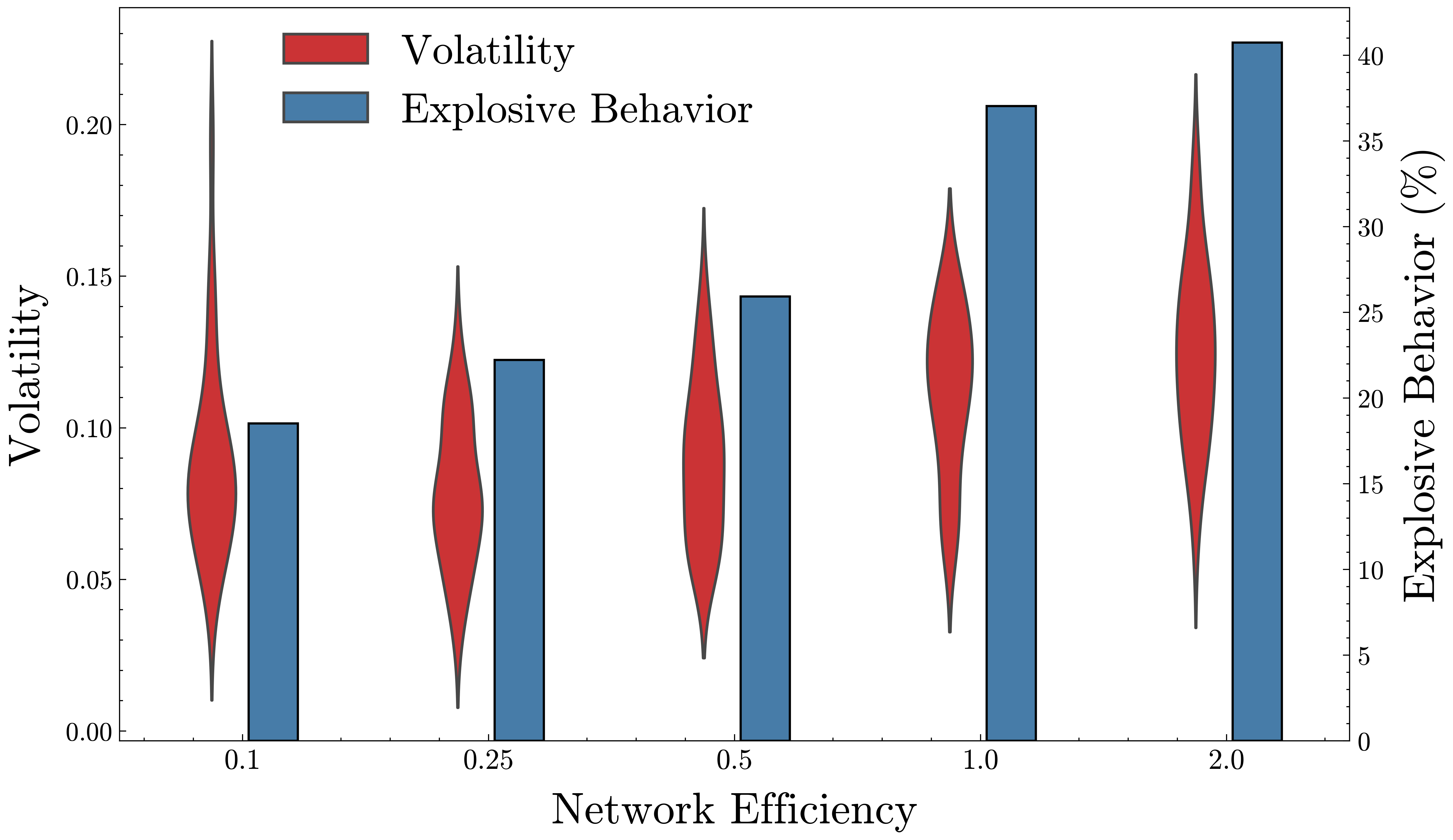}
\caption{Parameter set IV. Volatility and explosive behaviour both increase with network efficiency, $\phi$.}
\label{fig:eff-pset4}
\end{figure}

\cref{fig:eff-pset4} plots volatility and percentage of runs displaying explosive behaviour for different values of $\phi$. Clearly, there is a trend towards higher volatility and likelihood of explosive behaviour as $\phi$ increases. In other words, as the network becomes more efficient, the market becomes more volatile and prone to bubble formation. This result matches observations of real-world markets, where social media can drive explosive behaviours of unprecedented scale in so-called {\em meme stocks} \cite[e.g., see][]{Klein-2022}. The result is also consistent with previous findings in the literature, which suggest that social media (i.e., an efficient network) is correlated with higher volatility, whilst traditional news media (i.e., a comparatively inefficient network) is related to reduced volatility levels \citep{jiao-2020}. 

\begin{figure}[tb]
\centering
  \includegraphics[width=0.9\linewidth]{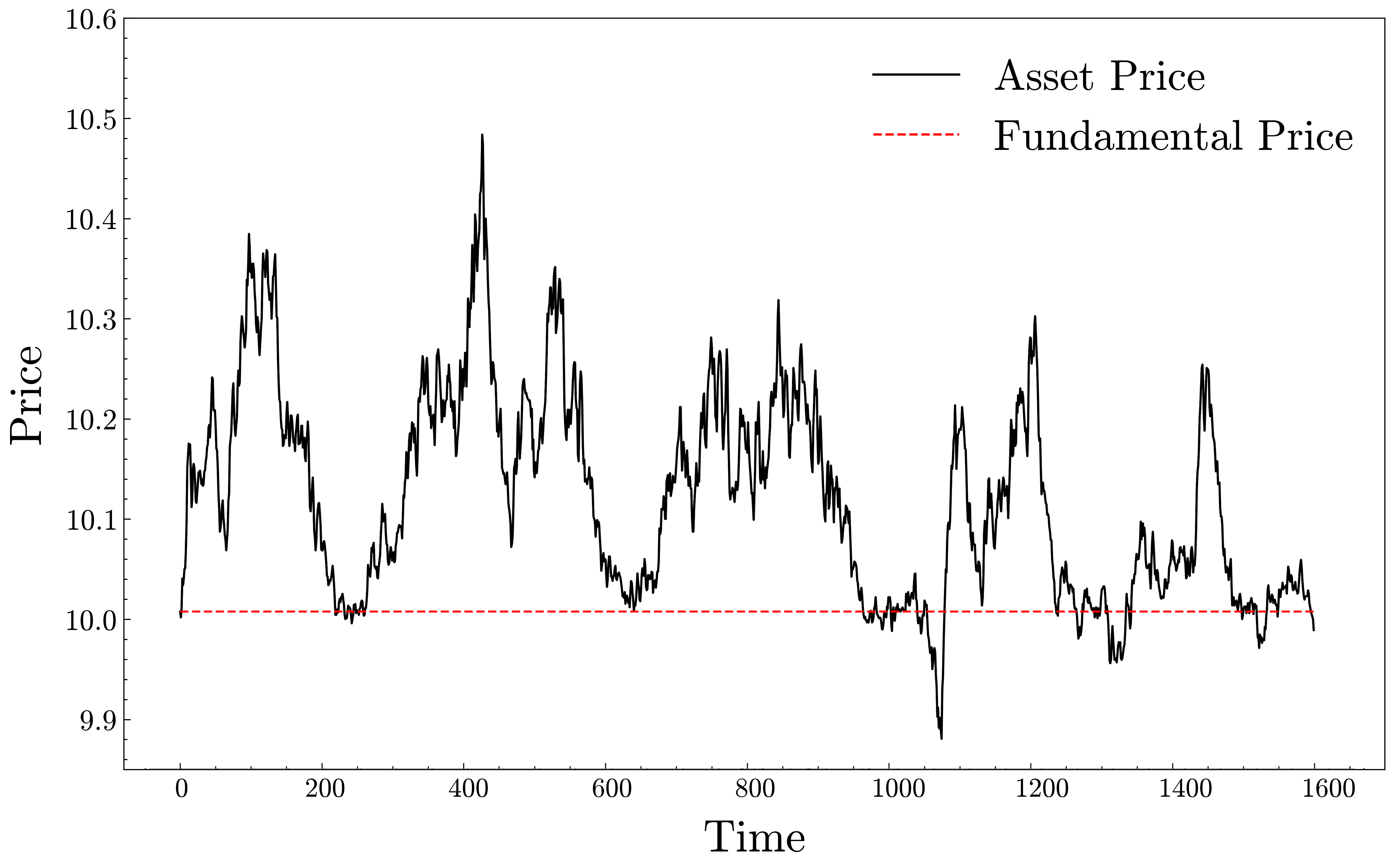}
\caption{Sample run of the asymmetric echo chamber scenario, showing a volatile asset price that remains consistently above the fundamental.}
\label{fig:example-bubbles-b2e2}
\end{figure}

\subsubsection{Echo chamber effect}
\label{section:echo}
\noindent
As described in \cref{sec:background-social-media}, social media can act as an {\em echo chamber}, which can lead to a strengthening of pre-existing opinions. We can simulate this effect by modifying $\omega$ and $\upsilon$ in equation \eqref{eq:community-nodes} via a third variable $\mathcal{E}$, such that the influence of a trader's opinion is multiplied by $\mathcal{E}$ if the trader's opinion conforms to the majority opinion within its parent community. We consider two distinct scenarios: 
\begin{itemize}
    \item \textbf{Asymmetric Model:} Where only optimists are affected by $\mathcal{E}$, such that: 
    \begin{equation}
    \label{eq:assymetric}
    \begin{aligned}
        \omega & = \begin{cases}
            \mathcal{E} & \text{if optimists $>$ pessimists}\\
            1 & \text{otherwise}
        \end{cases} 
    \end{aligned}
    \end{equation}
    \item \textbf{Symmetric Model:} Where all chartists are affected by $\mathcal{E}$, such that, in addition to the effect from equation \eqref{eq:assymetric}: 
    \begin{equation}
    \begin{aligned}
        \upsilon & = \begin{cases}
            \mathcal{E} & \text{if pessimists $>$ optimists}\\
            1 & \text{otherwise}
        \end{cases} 
    \end{aligned}
    \end{equation}
\end{itemize}

\begin{figure}[tb]
\centering
  \includegraphics[width=0.9\linewidth]{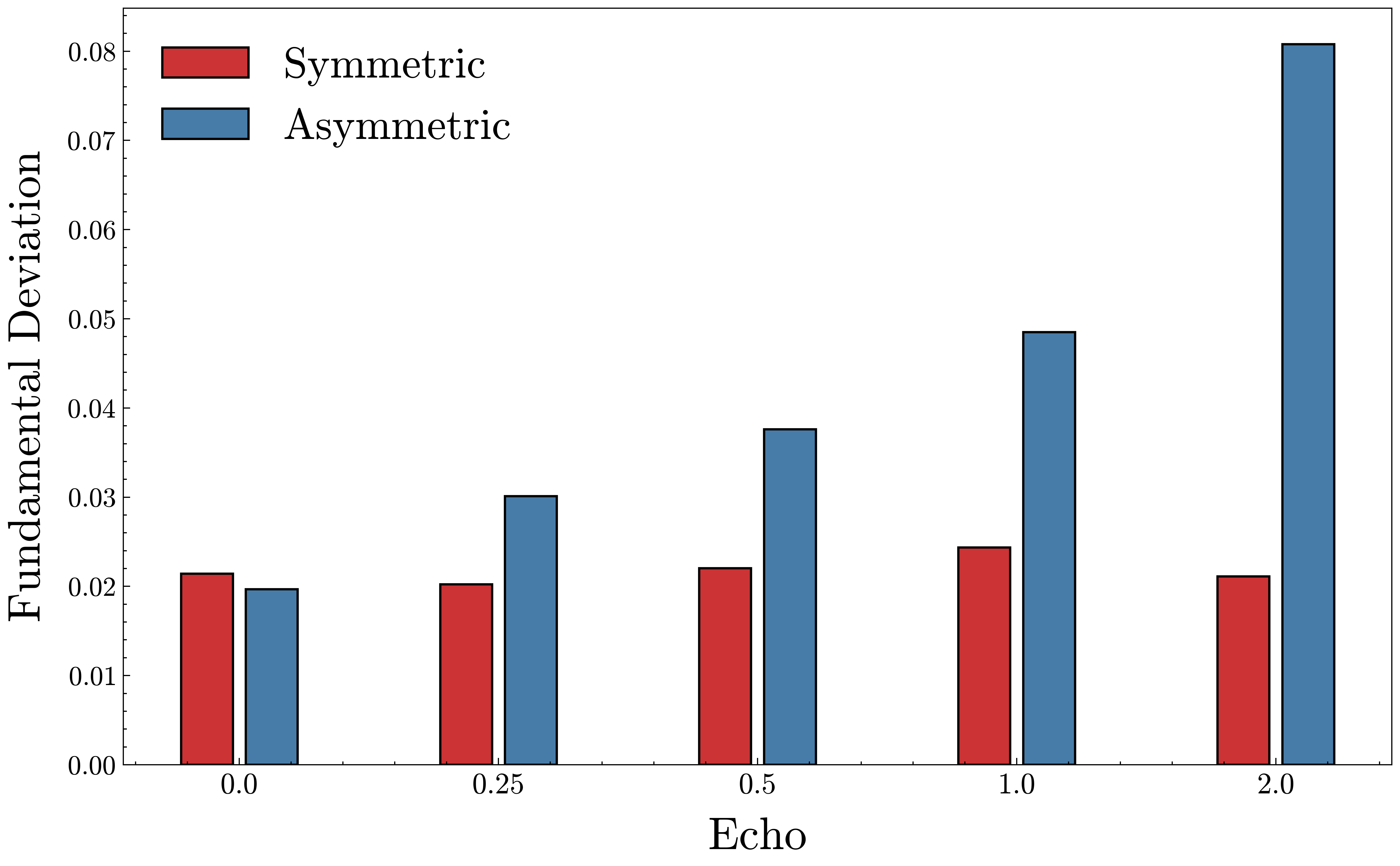}
\caption{Symmetric vs asymmetric echo chambers. Note that the asymmetric model exhibits a strong correlation between hierarchy strength and fundamental deviation, a relation not present for the symmetric model.}
\label{fig:sym-asym-fund}
\end{figure}

For small (but non-zero) values of $b$, increasing $\mathcal{E}$ in both the symmetric and asymmetric echo chamber scenarios has a similar effect to that of simply increasing the hierarchy strength.
However, for large values of $b$, these two scenarios differ significantly. Namely, whilst the symmetric scenario's effect on market behaviour becomes negligible, the asymmetric scenario is associated with large and persistent deviations between the asset's market price and its fundamental value (see \cref{fig:example-bubbles-b2e2,,fig:sym-asym-fund}). This result is consistent with the finding that echo chambers lead to higher levels of speculation, and this effect is exacerbated when it is asymmetric in the positive direction \citep{cookson-2022}. 

\begin{figure}[tb]
\centering
  \includegraphics[width=0.9\linewidth]{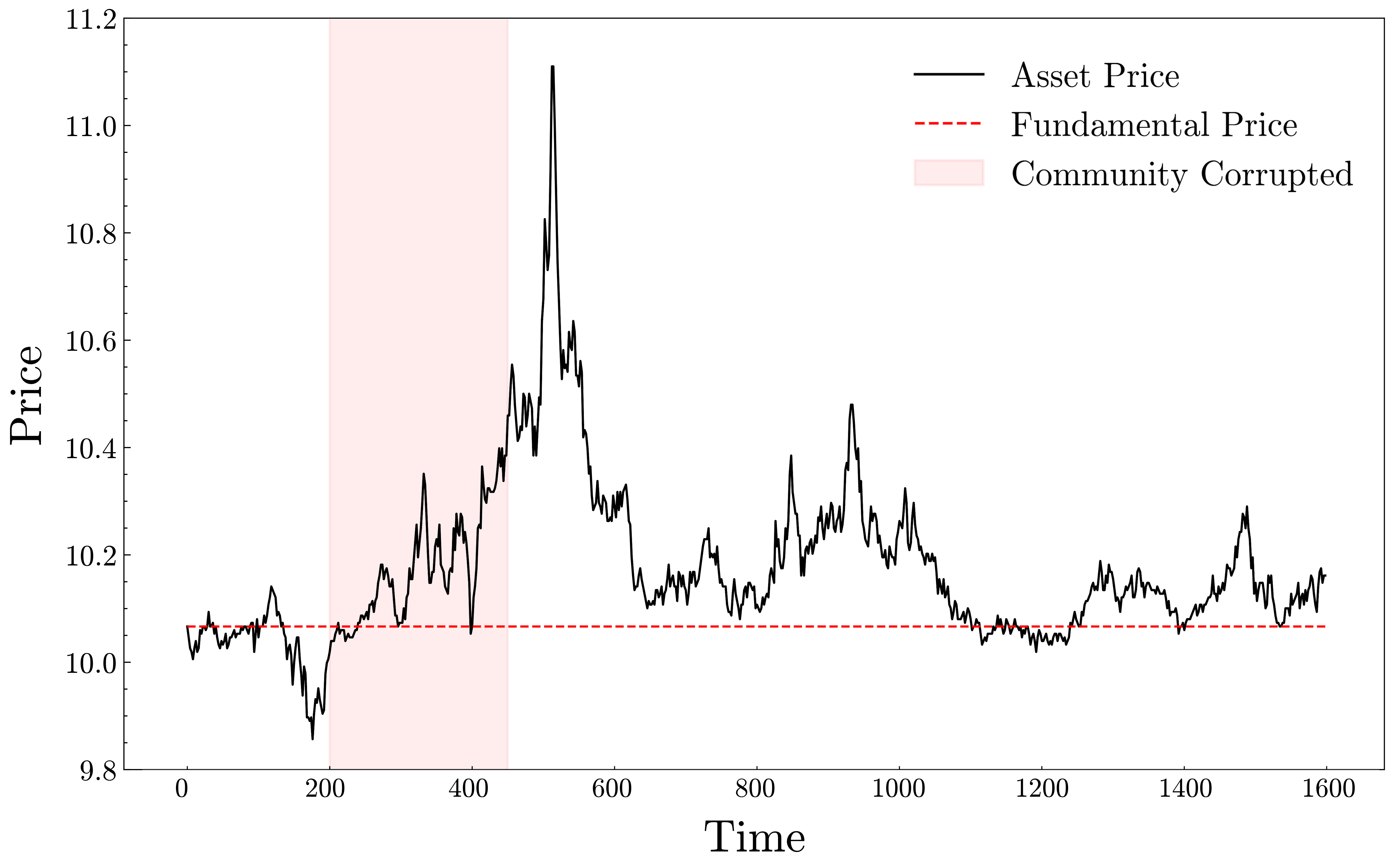}
\caption{Sample run for a successful pump-and-dump scheme where a community becoming corrupted (shaded red) results in a very pronounced peak in the asset price. Notice that the peak (T=550) occurs after the community ceases to be corrupted (T=450).}
\label{fig:example-pnd}
\end{figure}

\subsubsection{Pump-and-dump}
\label{section:pump}
\indent
We model pump-and-dump schemes through the process of a community becoming ``corrupted'': such that between time-steps $T_0$ and $T_1$, a corrupted community's (forward pass) signal conveys a deceptively high proportion of optimists to its children.
Hence, if a community is subject to becoming corrupted and the current time-step $t$ is between $T_0$ and $T_1$, its forward pass equation \eqref{eq:forward-pass} is replaced by equation \eqref{eq:forward-pass-corrupted}:
\begin{equation}\label{eq:forward-pass-corrupted}
C'_{[o,\ p,\ f]} = \frac{1}{2} C_{[o,\ p,\ f]} + \phi \cdot Q_{[S * (o + p + f),\ p,\ f]}
\end{equation}
\noindent where $S$ represents the \textit{signal strength} and $S * (o + p + f)$ is the magnitude of the signal emitted by the corrupted community.

We define a pump-and-dump scheme as being ``successful'' if the maximum price reached during or after the community is corrupted is larger than the maximum price reached by 95\% of 50 ``uncorrupted'' simulations using identical model parameter settings. \cref{fig:example-pnd} shows an example successful pump-and-dump scheme.
However, for all standard parameter sets (see \cref{tab:params}), the success rate of pump-and-dump schemes (with reasonable signal strengths s.t. $S \leq 50$) is very low, and not significantly greater than random chance. This should not come as a surprise, since pump-and-dump schemes are relatively infrequent, especially in mature markets predominated by professional investors who carefully monitor the fundamental value of assets.

Notwithstanding, some parameter configurations do lead to more frequently successful pump-and-dump schemes. Broadly, there are four distinct classes:
\begin{itemize}
    \item \textbf{Speculative markets:} where, by altering the transition pressures from equations \eqref{eq:U1} and \eqref{eq:optimist-pessimist-transition-modified}, the importance that traders place on the actual price trend of the asset, $\alpha_2$, or the profit made by other traders, $\alpha_3$, becomes much lower than the importance placed on one another's type, $b$. 
    \item \textbf{Volatile markets:} where the price changes quickly. This effect can be further exacerbated through changes to the parameters in equation \eqref{eq:total-demand}, such that the price becomes heavily influenced by the actions of speculative traders (i.e., large $t_c$), or only mildly influenced by fundamentalist traders (i.e., small $\gamma$).
    \item \textbf{Chartist-heavy markets:} where, through equation \eqref{eq:ep}, small values of $s$ make it unlikely for chartists to transition into fundamentalists. 
    \item \textbf{Nascent markets:} pump-and-dump schemes are generally more successful when $T_0$ is small, since the market is young and presumably still unstable.
\end{itemize}

In other words, markets that are speculative, volatile, nascent, and chartist-dominated are most susceptible to pump-and-dump. This fits with the common understanding of social media-driven pump-and-dump schemes, where targets are often na\"ive traders who are tricked into buying relatively obscure cryptocurrency assets after other users in a {\em WhatsApp} group claim to be making great profits.

\section{Discussion}\label{sec:discuss}
\noindent
We have introduced a new agent-based model of financial markets, where agent behaviours and opinions are influenced by a hierarchical network of communities. Our model extends the well-regarded Lux-Marchesi model (LM99) of financial markets, with the addition of a hierarchical influence network to simulate social media effects.  

We explored the general behaviour of the model and showed that it reproduces several stylized facts of real markets (\cref{tab:stylized-standard}; \cref{sec:results-stylised}), which strongly evidences model {\em realism}. Of particular note are the much higher autocorrelation of returns than the base model (LM99 logarithmic autocorrelation at $\tau = 10, T=70$ is smaller than 0.25 for all parameter sets, and squared autocorrelation is smaller than 0.2 \citep{gonzalo-MEng}). This higher degree of volatility clustering indicates a lower degree of market efficiency \citep{hameed-2006}, which is another sign pointing toward the model's validity, since strong social media effects are commonly believed to reduce the efficiency of the market upon which they act \citep{al-2018, bundi-2019}. The finding that a stronger hierarchy results in higher levels of volatility and more frequent instances of explosive behaviour (\cref{tab:volatility-results}) is also consistent with results presented by \citet{alfarano-2011note}, who conclude that a hierarchical structure is strongly correlated with markets displaying erratic behaviour. 

The model was then used to explore scenarios of real-world markets (\cref{sec:results-scenarios}), with findings confirming evidence in the literature, including: social media networks, which increase communication efficiency, leading to more volatile markets; echo chambers found in social media leading to higher levels of speculation; and pump-and-dump schemes being mostly successful in speculative, volatile, and chartist-dominated markets. These findings provide strong evidence of the validity and versatility of the model to explore a range of phenomena of financial markets. 

{\bf Related work:}
As described in \cref{sec:gap}, our model is most closely related to the MV23 model \citep{meine-2023}, which also extends LM99 with a hierarchical influence network. However, there are some key differences. (i) Our model's generalised network representation, which includes LM99 under some parameter configurations, allows for the gradual tuning and exploration of community effects. In contrast, MV23 presents a markedly more discrete jump in behaviour. (ii) With much simpler rules for agent behaviour transitions, our model can be more easily tailored for specific scenarios such as echo chambers and pump-and-dump schemes (see \cref{sec:results-scenarios}).

\section{Conclusions}\label{sec:conc}

We have extended a well-established financial ABM \citep{lux-marchesi-1999} to introduce a new hierarchical model of social media-driven influence on trading behaviours in financial markers. Empirical evidence shows that the model can succinctly and accurately replicate several emergent phenomena of real financial markets, especially those strongly influenced by social media.
To the authors' best knowledge, no other model in the literature has been shown to reproduce all these effects of social media, including the most similar model of \citet{meine-2023}.

However, our model is limited by some restrictive assumptions, such as: only a single asset is traded; every trader is affected by community influences equally; and the number of traders in the market is a fixed constant. In future works, we will relax some of these assumptions. A multi-asset market will be addressed following the approach of \citet{zhang-2017}, where each asset is modelled as a ``topic,'' and each individual's interest and opinion varies depending on the topic (i.e., asset); factors that affect speculative traders will be drawn from distributions that vary between traders; and traders will switch between an {\em active} and {\em inactive} state, allowing trade volumes to vary over time.
We will also focus on better understanding the causal relationships that lead to emergent phenomena, as well as exploring the simulation of other scenarios through the basic hierarchical structure presented in this paper.

\section{Funding}
This work was supported by UKRI grant \href{https://gow.epsrc.ukri.org/NGBOViewGrant.aspx?GrantRef=EP/Y028392/1}{EP/Y028392/1} ``AI for Collective Intelligence Research Hub''.


\balance


\bibliography{paper-refs}


\end{document}